# Experimental Demonstration of Efficient Spin-Orbit Torque Switching of an MTJ with sub-100 ns Pulses

Tanay A. Gosavi[1], Sasikanth Manipatruni[2], Sriharsha V. Aradhya[1], Graham E. Rowlands[1], Dmitri Nikonov[2], Ian A. Young[2], and Sunil A. Bhave[3]

[1] *Cornell University*
[2] *Exploratory Integrated Circuits, Components Research, Intel Corp., Hillsboro, OR 97124*
[3] *Purdue University*

Efficient generation of spin currents from charge currents is of high importance for memory and logic applications of spintronics. In particular, generation of spin currents from charge currents in high spin-orbit coupling metals has the potential to provide a scalable solution for embedded memory. We demonstrate a net reduction in critical charge current for spin torque driven magnetization reversal via using spin-orbit mediated spin current generation. We scaled the dimensions of the spin-orbit electrode to 400 nm and the nanomagnet to 270 nm × 68 nm in a three terminal spin-orbit torque, magnetic tunnel junction (SOT-MTJ) geometry. Our estimated effective spin Hall angle is 0.15-0.20 using the ratio of zero temperature critical current from spin Hall switching and estimated spin current density for switching the magnet. We show bidirectional transient switching using spin-orbit generated spin torque at 100 ns switching pulses reliably followed by transient read operations. We finally compare the static and dynamic response of the SOT-MTJ with transient spin circuit modeling showing the performance of scaled SOT-MTJs to enable nanosecond class non-volatile MTJs.

AUTHOR EMAIL ADDRESS: bhave@purdue.edu

# Introduction

Continued scaling of electronic devices has enabled unprecedented growth of computing performance in the past few decades. However, as the electronic device gate lengths approach 20 nm dimensions [1], further scaling of complementary metal oxide semiconductor (CMOS) field-effect transistors (FET) require the realization of several complex material and device changes [2]. One of the methods for enabling further scaling in computing performance is by integration of spintronic devices for on-chip embedded memory [3] or spintronic logic [4]. An efficient method of switching magnetic bits is a pre-requisite to enable such integration [4]. In this context, spin transfer torque (STT) switching of nanomagnets driven by CMOS transistors [5] has attracted significant effort. However, traditional STT devices are reliant on high spin polarization in magnetic tunnel junctions (MTJ) due to orbital symmetry dependent spin filtering in oxides (e.g. MgO) [6]. MTJ based MRAM technology has the following fundamental drawbacks for compatibility with future scaled CMOS technologies: a) Incompatibility of the high operating voltages and currents required for MTJ tunnel currents with scaled CMOS. b) Large access transistor size required to meet the drive current requirement limits circuit density. c) Reliability issues created by the high voltage (>0.8V) and high current (>100 µA) damaging MTJ oxide. Hence it is of great interest to pursue MTJ switching methods alternative to STT, which would provide high spin polarization at low voltage and low current operation.

The discovery of Giant Spin-orbit Torques (GSOT) (Giant Spin Hall Effect (GSHE) and Rashba Effect (RE) [7-10]), in which large spin polarized currents are generated transverse to the charge current direction in high-Z metals (such as Pt [11-13, 8-9], β-Ta [9], β-W [10], doped Cu [e.g. 14-15]), may provide a solution to the voltage, current scaling and reliability problems of magnetic embedded memory. In particular, scaled SOT-MTJs can have a) better drivability in scaled CMOS; b) fast switching time approaching 500 ps; c) decoupled read and write paths; d) improved trade-off of non-volatility vs. write time [16].

**Description of the device**

We fabricated a three terminal device which forms the basic cell for the MRAM which has a spin orbit torque induced write mechanism and MTJ based read-out as shown in Figure 1, using fabrication technique shown by Aradhya et al., in [28]. The device material stack comprising substrate/Ta(6)/$Co_{40}Fe_{40}B_{20}$(1.6)/MgO(1.6)/$Co_{40}Fe_{40}B_{20}$(4)/Ta(3)/Ru(3)/Ti(6)/Pt(40) (thicknesses $t$ are in nanometers) is patterned into the geometry as shown in Figure 1A. The device has narrow rectangular channel of width $W_{ch}$= 400 nm, 500 nm, 600 nm and length $L_{ch}$=1600, 2000 and 2400 nm made of β-Tantalum (β-Ta) which couples via the spin orbit torques to $Co_{40}Fe_{40}B_{20}$ free layer of the MTJ. The small thickness of the free layer is chosen to reduce the demagnetization field and hence the switching energy of the nanomagnet, but it is thick enough to keep the magnetization in the plane of the device. Isometric view of the SHE-MTJ showing the top electrode and bottom electrode are shown in Fig. 1B. Top-view is shown in Fig. 1C where the nanomagnet is oriented along the width of the GSOT electrode for appropriate spin injection. Ti/Pt contacts are made to the spin orbit electrode. This MTJ part of dimension 270nm × 68nm was fabricated using E-beam lithography on top of the β-Ta channel.

The top leads of Ti/Pt bilayer were added on the Ta/Ru stack to make contacts to the MTJ. The magnetization state of the cell is written by applying a charge current via the GSOT electrode. The direction of the magnetic writing is set by the direction of the applied charge current. Positive currents (along +y direction) produce a spin injection current with transport direction (+z) and spins polarized in (+x) direction. The injected spin current in its turn produces spin torque to align the nanomagnet in the (-x) direction (since the spin hall angle for β-Ta is negative). If the direction of the current is switched to opposite, so is the magnetization direction. The transverse spin current ($\vec{I}_s = \vec{I}_\uparrow - \vec{I}_\downarrow$ with spin direction $\hat{\sigma}$) for a charge current ($\vec{I}_c$) in the write electrode is given by

$$\vec{I}_s = P_{SHE}(w, t, \lambda_{sf}, \theta_{SHE})(\hat{z} \times \vec{I}_c) \tag{1}$$

Where $P_{SHE}$ is the spin Hall injection efficiency which is the ratio of magnitude of transverse spin current to lateral charge current, $w$ is the width of the nanomagnet, $t$ is the thickness of the GSHE metal electrode, $\lambda_{sf}$ is the spin flip length in the GSHE metal, $\theta_{SHE}$ is the spin Hall angle (coefficient) for the GSHE-metal to the FM1 interface.

We demonstrate bidirectional switching of the three terminal MTJ using spin injection from the spin orbit write electrode. The switching diagram of the SHE-MTJ with an external magnetic field is shown in Figure 2A. The small signal resistance (dv/di) is plotted as a function of applied magnetic field, showing a clear hysteresis window with single domain switching of the free layer. The major loop is shown in the inset. The bidirectional current induced switching is shown in Figure 2B in the presence of an offset field of 102 Oe to center the hysteresis curve for the device. We verify the symmetry of the switching by using a negative offset field (-102 Oe). For positive offset field, the MTJ switches to AP/P when the current through GSOT is positive/negative. When the off-set field is reversed, the MTJ switches to P/AP when the current through GSOT is negative/positive. In an integrated device with a synthetic anti-ferromagnet, the offset field is not necessary and the relationship of the switching and current direction is purely given by the equation 1.

We show the effect of scaling the SOT electrode on the switching of the SOT-MTJ by comparing the DC switching properties for a 270 nm × 68 nm magnet, with SOT electrode of various width: 400 nm, 500 nm, and 600 nm. In Figure 3A, bidirectional spin torque switching of the MTJs as a function of applied current in the SOT electrode is plotted. The critical currents for parallel-to-antiparallel switching and vice versa, $I_{cP-AP}$ and $I_{cAP-P}$, are comparable showing that symmetric bidirectional switching is feasible with SOT switching. Symmetric switching can simplify the constraints on the driving circuitry used for MRAM.

We extract the zero temperature critical currents and magnetic energy barriers for the devices using ramp rate measurements. Figure 3B, shows the critical current as a function of ramp rate $\dot{I}_c$. Using thermal activation model, this dependence can be fit to a function form $I_c = I_{c0}\left[1 - \frac{k_B T}{U}\ln(\frac{k_B T}{U} \cdot \frac{I_{c0}/\tau_0}{\dot{I}_c})\right]$, where the energy barrier U and the zero-temperature critical current $I_{c0}$ are the fitting variables, $k_B T$ is the thermal energy. We assume that the temperature of the device was close to room temperature, and the attempt time $\tau_0 \sim 1 ns$. The extracted barriers and zero temperature critical current are shown in Figure 3D. The ratio of the critical current at zero temperature to the energy barrier is plotted in Figure 3C showing linear scaling.

We extract the 'effective spin Hall angle' ($\theta'_{SHE}$) for the device by taking the ratio of the expected critical spin current density in spin transfer torque switching ($J_{s0-MTJ}$) to the charge current density $J_{c0-SHE}$ derived from the critical current in spin Hall switching $I_{c0-SHE}$.

$$\theta'_{SHE} = \frac{J_{s0-MTJ}}{J_{c0-SHE}} = (W_{ch}t_{ch})\frac{J_{s0-MTJ}}{I_{c0-SHE}} = \frac{(W_{ch}t_{ch})}{I_{c0-SHE}}\left(\frac{2e}{\hbar}\mu_0 M_s t'_m \alpha\left(H_k + \frac{M_{eff}}{2}\right)\right) \quad (2)$$

The calculated spin Hall angles for the three devices are 0.16 ($W_{ch}$ = 400 nm), 0.18 ($W_{ch}$ = 500 nm) and 0.20 ($W_{ch}$ = 600 nm). Here the magnet thickness without the dead layer is $t'_m$, the saturation magnetization $M_s$, the effective demagnetization is $M_{eff}$, the material anisotropy effective field is $H_k$, and Gilbert damping $\alpha$. We assumed a dead layer of 0.2 nm. This technique for effective spin Hall angle extraction is approximate due to the following factors: a) the approximate estimate of the dead layer thickness ($t_m$-$t'_m$); b) $I_{c0\text{-SHE}}$ is not exactly proportional to $W_{ch}$ showing a large offset; c) ignoring the contribution of the field like torque arising due to reflection of spins at the interface and due to the interface Rashba effect.

We also calculate the effective spin injection efficiency in the measured spin orbit effect device:

$$P_{eff} = \frac{I_{s0-MTJ}}{I_{c0-SHE}} = \frac{\pi ab J_{s0-MTJ}}{4 I_{c0-SHE}} \approx 96 - 113\% \tag{3}$$

and show that the critical current in SHE is ~2x smaller the critical current in STT with MTJ spin polarized $P_{MTJ}$ currents:

$$\frac{I_{C0-MTJ}}{I_{c0-SHE}} = \frac{P_{SHE}}{P_{MTJ}} \sim 2 \tag{4}$$

The above effective spin Hall angle is calculated under the assumption that the spin torque is only due to Slonczewski spin torque.

We obtain intrinsic bulk spin Hall angle to be 0.35-0.44 by accounting for the finite interface conductivity, the interface transmission coefficient, and spin diffusion from the second surface of β-Ta which suppress the spin injection into the magnet. The effect of finite interface reflectivity (i.e. the imaginary component of the mixing conductance ($\tilde{G}^{\uparrow\downarrow}$) and spin diffusion from bottom interface [25] under a semi-classical spin diffusion equation is given by

$$\theta'_{SHE}(t) = \theta_{SHE} \frac{\left(1-e^{-t/\lambda_{sf}}\right)^2}{\left(1+e^{-2t/\lambda_{sf}}\right)} \times \left[ \frac{\left|\tilde{G}^{\uparrow\downarrow}\right|^2 + \text{Re}\left[\tilde{G}^{\uparrow\downarrow}\right]\tanh^2(t/\lambda_{sf})}{\left|\tilde{G}^{\uparrow\downarrow}\right|^2 + 2\text{Re}\left[\tilde{G}^{\uparrow\downarrow}\right]\tanh^2(t/\lambda_{sf}) + \tanh^4(t/\lambda_{sf})} \right] \tag{5}$$

Where $\tilde{G}^{\uparrow\downarrow}(t) = G_{mix} 2\rho_{Ta} \lambda_{sf} \tanh(t/\lambda_{sf})$ is a scaled spin mixing conductance accounting for thickness (t) induced effect, $\rho_{Ta}$ is the resistivity, $\lambda_{sf}$ is the spin-flip length in Ta [26].

## Transient switching of the device and comparison to spin circuit model:

We show the operation of the three-terminal SOT device as a memory element by performing consecutive write and read cycles. The write cycles are performed at 4X the critical current (we note that such high biasing is not possible with tunnel junction based spin torque switching) followed by a read pulse of amplitude 50 mV applied to the third terminal of the MTJ. The repeatable bi-directional transient switching current and voltage data are plotted in figure 4. Figure 4A shows the switching of the device with 10 us pulse width write pulses followed by a read pulse of 10 us width with ~ 100 nA of write currents. Figure 4B, 4C, 4D show the read and write of the SOT MTJ with progressively lower pulse widths down to 100 ns. The pulse shape for 100 ns switching is shown in in inset of figure 4D.

## Vector spin circuit modeling of spin-orbit torque switching devices

We model the spin-orbit torque MTJ using vector spin circuit theory [17] comprising 4x4 conduction matrix formulation for spin transport coupled with magnetization dynamics. The circuit model is shown in figure 5. The proposed model can be self-consistently coupled to the nanomagnet dynamics including the thermal stochastic noise effects [18]. The spin torque acting on the free layer in a spin-orbit torque MTJ originates from a) spin torque due to spin injection from the fixed layer b) spin torque due to the spin orbit torque acting on the free layer. The phenomenological equation describing the dynamics of nanomagnet with the magnetic moment unit vector ($m$) is the modified Landau-Lifshitz-Gilbert-Slonczewski (LLG) equation [19]:

$$\frac{\partial m}{\partial t} = -\gamma \mu_0 [m \times \bar{H}_{eff}] + \alpha \left[ m \times \frac{\partial m}{\partial t} \right] + \frac{1}{eN_s} \bar{I}_{s\perp m}(V,G) \qquad (6)$$

where $\gamma$ is the electron gyromagnetic ratio; $\mu_0$ is the free space permeability; $\vec{H}_{eff}$(T) is the effective magnetic field due to material, shape, and surface anisotropies, with the thermal noise component [18] and $\vec{I}_{s\perp} = (\hat{m} \times \hat{m} \times \vec{I_s})$ is the component of vector spin current perpendicular to the magnetization, $N_s$ is the total number of Bohr magnetons in the magnet. The dynamics of the MTJ are solved self-consistently with the spin transport in the equivalent circuit models.

The equivalent vector spin circuit for SOT-MTJ comprises of the equivalent spin conductances of the fixed Ferromagnet (FM$_{fix}$) and free Ferromagnet (FM$_{free}$) interfaces to form the MTJ [21]. The vector spin equivalent circuit model for an SOT-MTJ is described in figure 5A. The model comprises of three nodes N0, N1 and N2 to describe the MTJ. The magnetization of top fixed layer and bottom free layer are described by $\hat{m}_{fix}$ and $\hat{m}_{free}$. The 4-component conductivity of the FM1 and oxide interface is described by G$_{FM1}$ and conductivity of the FM2 and oxide interface is described by G$_{FM2}$. The conductance matrix describing the spin transport across a FM/Oxide interface can be written as:

$$\begin{bmatrix} I_c \\ I_{sx} \\ I_{sy} \\ I_{sz} \end{bmatrix} = \begin{bmatrix} G_{11} & \alpha(V_c)G_{11} & 0 & 0 \\ \alpha(V_c)G_{11} & G_{11} & 0 & 0 \\ 0 & 0 & G_{SL}(V_c) & G_{FL}(V_c) \\ 0 & 0 & -G_{FL}(V_c) & G_{SL}(V_c) \end{bmatrix} \begin{bmatrix} V_c \\ V_{sx} \\ V_{sy} \\ V_{sz} \end{bmatrix}$$

(7)

Where G$_{11}$ is the interface conductivity (per interface) of the FM/MgO interface, $\alpha(V)$ is the spin polarization across the interface as a function of voltage, G$_{SL}$(V$_c$) and G$_{FL}$(V$_c$) are Slonczewski and field like torque contributions from the tunneling spin current across the interface. The voltage dependence of spin polarization $\alpha(V)$, G$_{SL}$(V), G$_{FL}$(V) is dependent on the detailed band structure of the electrodes and tunneling materials [22-24]. The effect of magnetization rotation for a precessing MTJ can be described using the proposed model, where the 4 component conductances evolve as a function of the magnetization of the free magnet.

$$G_{FM0}(\hat{m}) = R(\hat{m})^{-1} G_{FM0}(\hat{x}) R(\hat{m}) \qquad (8)$$

Where R is a 4-component transformation to rotate the conductance matrices.

The spin torque from tunneling spin currents acting on the magnet and the effect of spin torque from spin orbit layer are included via a spin injection into the free layer as governed by the physics of spin injection from SOT layer to FM [25-26]. The equivalent spin circuit model includes a current control spin current to model the injection of spin current from the SOT electrode to the free layer. We also include the field like component of the SOT via a current controlled effective magnetic field due to the SOT [27]. A coupled simulation of the spin torque dynamics of an MTJ driven by the spin current response from a vector spin circuit model is shown in Figure 5A. The vector magnetization dynamics of the free layer are shown in figure 5B, showing magnetic switching under the combined action of anti-damping spin torque and effective field due to spin orbit effects. The circuit simulation captures the change in the state of the MTJ via a 10 mV relative voltage bias across the device. The change in the MTJ resistance is shown in figure 5B. Vector spin equivalent circuit is shown in figure 5C. Switching time vs. applied voltage pulse characteristics are shown in Figure 5D. The combined effect of field-like spin torque and anti-damping toque significantly enhance the speed of switching showing nominal switching speed of < 200 ps for a 70 kT magnet with dimensions of 20 X 60 nm with spin orbit metallic electrode of 60 X 60 nm of resistivity 200 µm.cm. We assumed a bulk spin hall ratio of 0.2 and effective Rashaba field of $4 \times 10^{-6}$ Oe/(A/cm$^2$) for the transient vector spin simulations.

## Conclusion

We have demonstrated transient switching and read out of magnetization in a spin orbital (spin Hall) effect MTJ. Scaled SOT metallic electrodes produce a favorable scaling of spin currents. We estimated a 2X improvement in spin current efficiency compare to tunneling spin transfer torque mechanism. Using zero temperature critical current extraction, we derive an effective spin hall angle of 0.2. Accounting for imperfect interfaces, we deduce the intrinsic spin hall angle to be 0.33-0.40. Using vector spin circuit modeling, we show that further aggressive scaling of dimensions can lead to nanosecond class spin orbit effect switching devices with non-volatility.

## Acknowledgments

We acknowledge Dr. S. Aradhya, Dr. G. E. Rowlands and M-H Nguyen of Cornell University for help with fabricating SOT-MTJ devices and with measurement setups. We also thank Prof. Daniel C. Ralph and Prof. Gregory D. Fuchs for discussions about devices and measurements. We are grateful to Dr. P. Gowtham for his encouragement throughout the project. This work was supported in part by the Semiconductor Research Corporation and the Cornell Center for Material Research with funding from NSF MRSEC program (DMR- 1120296). This work was performed in part at Cornell NanoScale Facility, a node of National Nanotechnology Infrastructure Network, which is supported by NSF (Grant ECCS-0335765).

induced spin currents by a Hf spacer layer in W/Hf/CoFeB/MgO layer structures. Applied Physics Letters, 104(8), 082407.

[28] Aradhya, S. V., Rowlands, G. E., Oh, J., Ralph, D.C., & Buhrman, R. A. (2016). Nanosecond-timescale low error switching of in-plane magnetic tunnel junctions through dynamic Oersted-field assisted spin-Hall effect. Nano Letters, 10.1021/acs.nanolett.6b01443

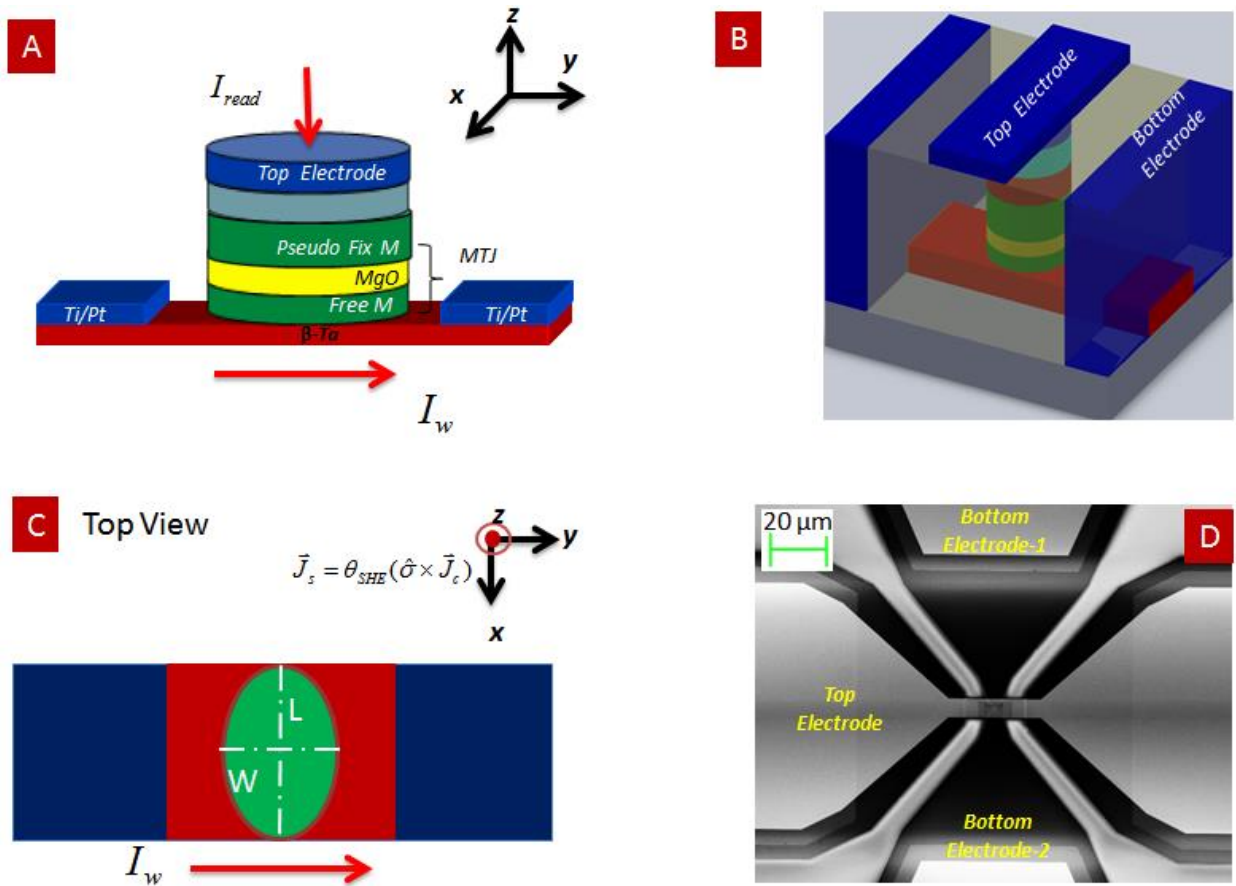

Figure 1 A) A Three Terminal Spin Hall Memory Device with spin orbit effect write electrode and MTJ based readout B) Isometric view of the bit showing the device C) Top view of the cell showing the orientation of the free layer magnet and the GSOT metal D) SEM image of the test structure.

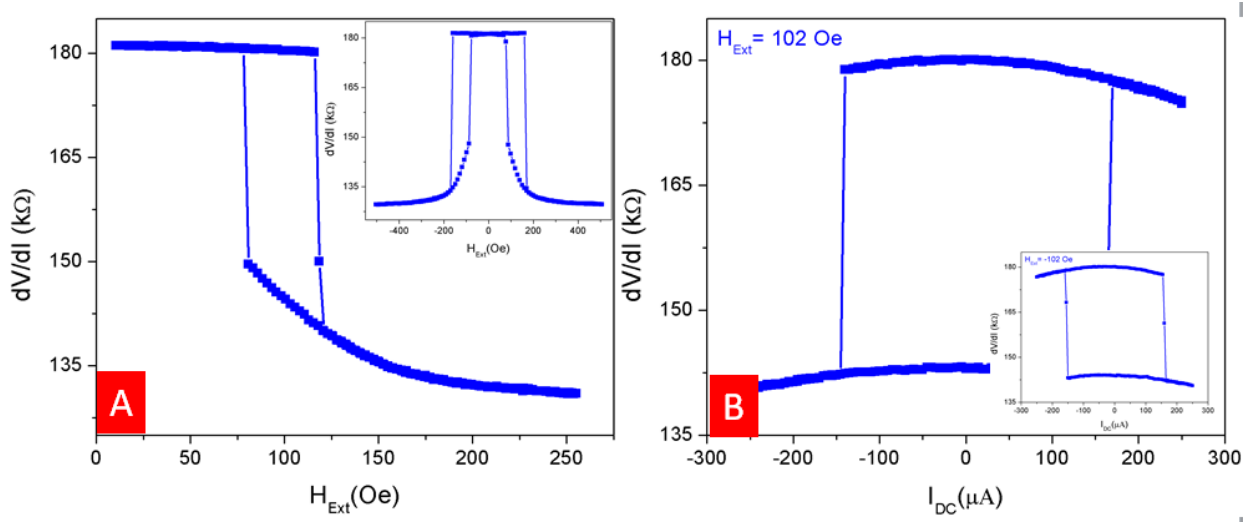

Figure 2: A) TMR Minor loop of the MTJ as a function of the external applied in-plane magnetic field $H_{Ext}$ along the long axis of the device. (Inset) TMR major loop of the device. B) TMR of the MTJ as a function of applied DC current $I_{DC}$. The measurement was performed with a bias in-plane magnetic field of 102 Oe to center the current loop.

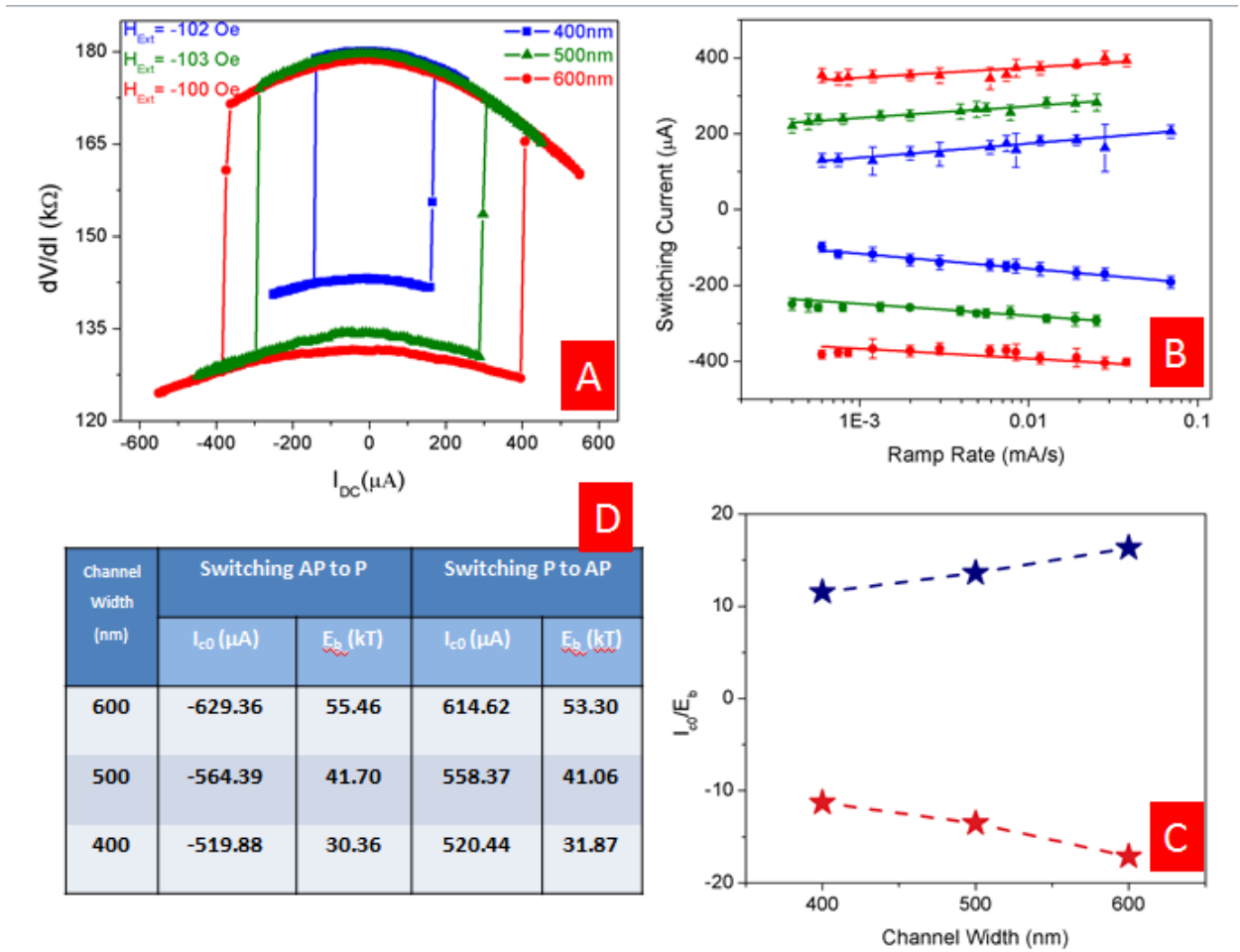

Figure 3. A) Effect of channel width reduction on identical magnets switched with spin current injection via SOT B) Switching currents as function of ramp rate for 270×68 nm$^2$ device fabricated on channels of width 600 nm (red), 500 nm (blue) and 400 nm (green). Currents for AP to P switching is shown using circles while P to AP switching is shown using squares. Solid lines represent linear fits of switching current versus ramp rate. Error bars are smaller than the symbol sizes. C) Linear scaling of critical current normalized by magnetic barrier with channel width for AP to P and P to AP switching D) Table showing values of zero-thermal-fluctuation switching current $I_{c0}$ and the energy barrier U against thermally activated magnetic reversal calculated from linear fits shown in Fig 3B.

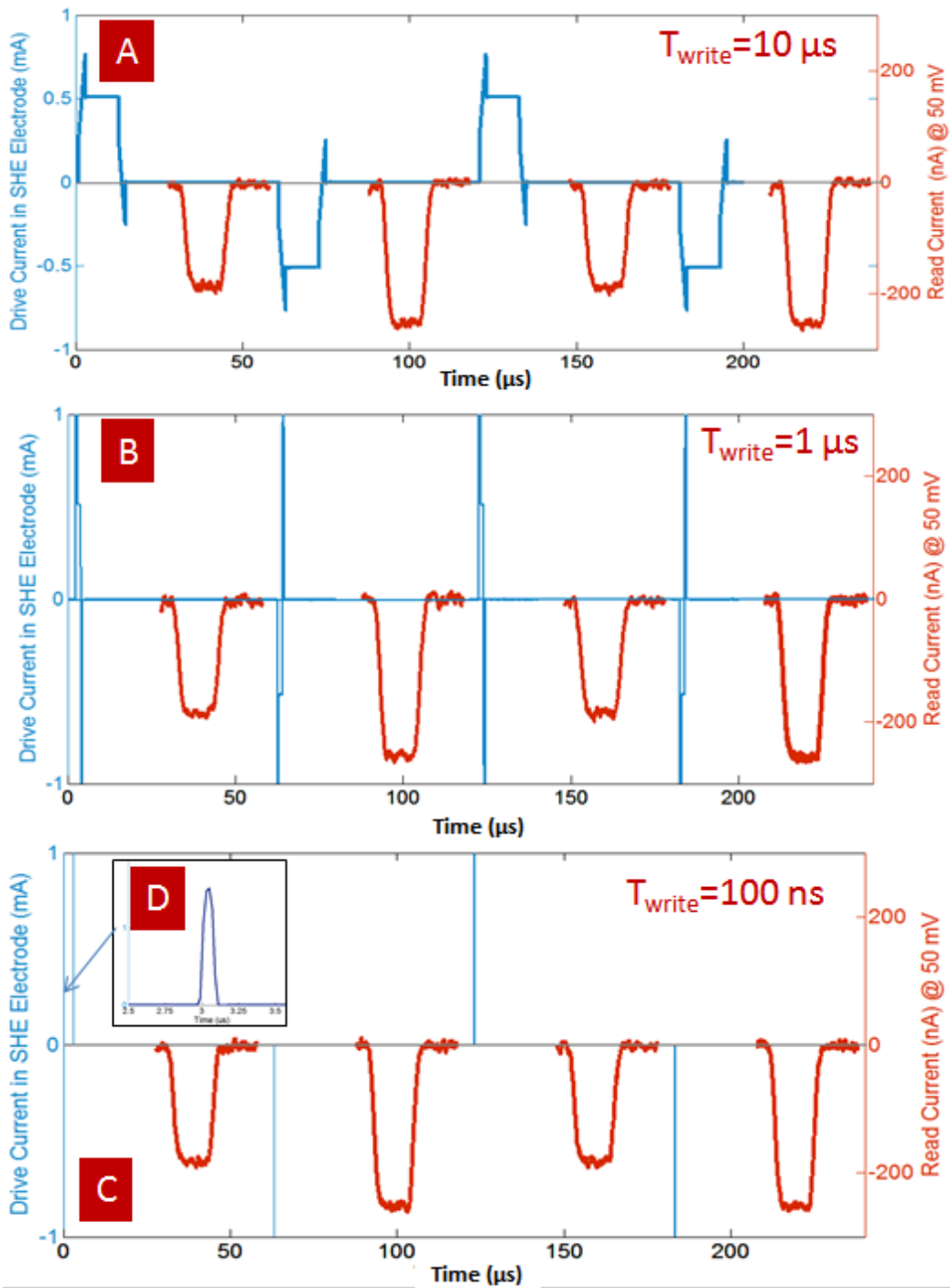

Figure 4. Transient write and reading of the SHE-MTJ with write pulse width a) 10 μs b) 1 μs c) 100 ns. D) inset showing the 100 ns width of write pulse. Read is performed with a 50 mV read voltage pulse applied between the write operations.

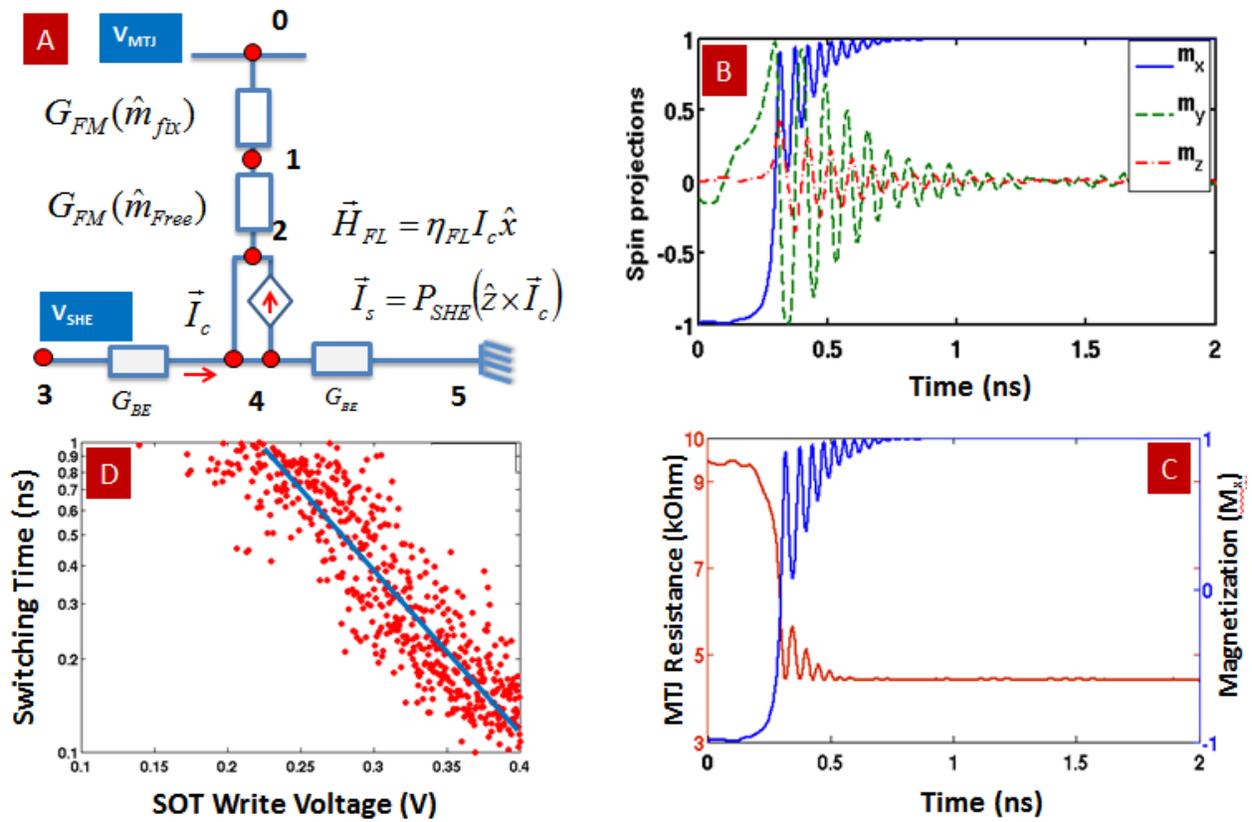

Figure 5. A) Transient magnetization switching of the SOT-MTJ nanomagnet B) Transient resistance switching of the SOT-MTJ C) Vector spin circuit model of SOT MTJ with anti-damping spin current and field-like torques D) Stochastic simulation of SOT-MTJ showing nominal switching at 0.4 V applied voltage.

19